\documentclass{article}

\usepackage[frozencache,cachedir=.]{minted}
\usepackage{graphicx}
\usepackage[alldates=year,backend=biber,maxnames=9,maxcitenames=2,doi=true,hyperref=true,giveninits=true,uniquename=false]{biblatex}
\usepackage{url}
\title{parasweep: A template-based utility for generating, dispatching, and post-processing of parameter sweeps\footnote{This preprint has been published as \fullcite{bach_parasweep_2021}}}
\author{Eviatar Bach\footnote{Department of Atmospheric and Oceanic Science and the Institute for Physical Science and Technology, University of Maryland, College Park, MD. E-mail: eviatarbach@protonmail.com.}}
\date{}

\renewbibmacro*{doi+eprint+url}{%
    \printfield{doi}%
    \newunit\newblock%
    \iftoggle{bbx:eprint}{%
        \usebibmacro{eprint}%
    }{}%
    \newunit\newblock%
    \iffieldundef{doi}{%
        \usebibmacro{url+urldate}}%
        {}%
    }

\DeclareSourcemap{
  \maps[datatype=bibtex]{
    \map{
      \step[fieldset=issn, null]
    }
  }
}

\DeclareSourcemap{
  \maps[datatype=bibtex]{
    \map{
      \step[fieldset=urldate, null]
    }
  }
}

\DeclareSourcemap{
  \maps[datatype=bibtex]{
    \map{
      \step[fieldset=pagetotal, null]
    }
  }
}

\AtEveryBibitem{%
  \clearlist{language}%
}

\begin{document}

\maketitle

\begin{abstract}
	We introduce parasweep, a free and open-source utility for facilitating parallel parameter sweeps with computational models. Instead of requiring parameters to be passed by command-line, which can be error-prone and time-consuming, parasweep leverages the model's existing configuration files using a template system, requiring minimal code changes. parasweep supports a variety different sweep types, generating parameter sets accordingly and dispatching a parallel job for each set, with support for local execution as well as common high-performance computing (HPC) job schedulers. Post-processing is facilitated by providing a mapping between the parameter sets and the simulations. We demonstrate the usage of parasweep with an example.
\end{abstract}

\section*{Keywords}

parameter sweeps; parallel computing; distributed computing; parametric modelling; Python; scientific computing

\begin{table}[!h]
\begin{tabular}{p{6.5cm}p{6.5cm}}
\textbf{Code metadata} & \textbf{} \\
\hline
Current code version & 2021.01 \\
Permanent link to code/repository used for this code version & \url{https://github.com/eviatarbach/parasweep} \\
Legal Code License   & MIT License \\
Code versioning system used & git \\
Software code languages, tools, and services used & Python \\
Compilation requirements, operating environments \& dependencies & xarray \cite{hoyer_xarray_2017} version 0.9+, NumPy \cite{van_der_walt_numpy_2011}, and SciPy \cite{jones_scipy_2001}. To use the optional DRMAA functionality, DRMAA Python, DRMAA, and a DRMAA-compatible job scheduler and its DRMAA interface are required. To use the optional advanced template language, Mako is required. \\
Link to developer documentation/manual & \url{http://www.parasweep.io/en/latest/} \\
Support email for questions & \url{eviatarbach@protonmail.com}\\
\hline
\end{tabular}
\end{table}

\begin{table}[!h]
\begin{tabular}{p{6.5cm}p{6.5cm}}
\textbf{Software metadata description} & \\
\hline
Current software version & 2021.01 \\
Permanent link to this version  & \url{https://github.com/eviatarbach/parasweep/releases/tag/2021.01} \\
Computing platforms/Operating systems & Operating systems with a Python interpreter (Linux, Microsoft Windows, and macOS, for example) \\
\hline
\end{tabular}
\end{table}

\section{Motivation and significance}

Parameter sweeps, whereby computational models are run repeatedly with different sets of parameters, are widely used in a plethora of scientific fields \parencite{wilkinson_grid_2009,sudholt_application_2005}. They can be done for a variety of reasons, such as testing sensitivity of a model to its parameters \parencite{edwards_uncertainties_2005}, exploring the qualitative changes in the behavior of a model as parameters are varied (for example, bifurcations) \parencite{marsh_bistability_2004}, or to find values of parameters that optimize some criterion \parencite{sudholt_application_2005}. The latter use is often employed for hyper-parameter optimization in machine learning \parencite{goodfellow_deep_2016}. Parameter sweeps are a classic example of an ``embarrassingly parallel'' problem, in that the set of simulations can easily be run in parallel because each simulation does not have to exchange information with the other simulations. However, most model software does not have built-in parameter sweep functionality that allows for generating parameter sets and running each instance in parallel.

We present parasweep, a free and open-source utility for easily carrying out parallel parameter sweeps for any computational model, with support for individual multi-core computers, clusters, and grids. It is written in Python, an interpreted, cross-platform language widely used for scientific applications; however, parasweep can work with models in any language. It makes use of configuration file templates in order to easily dispatch simulations with different parameter sets. The process of executing a parameter sweep and the full set of features of parasweep is discussed in section 2.

Previous papers have focused on how to efficiently allocate resources for large parameter sweeps on various infrastructures \cite{casanova_apples_2000,abramson_high_2000,wilson_launcher_2014}. parasweep does not incorporate any special scheduling strategies, but supports a number of cluster and grid schedulers through the Distributed Resource Management Application API (DRMAA), a standardized interface for communicating with job schedulers. Several tools have also been developed specifically for parameter sweep applications. One such tool, Nimrod \cite{abramson_high_2000}, is only available for grid systems. ILab \cite{yarrow_advanced_2000,devivo_comparison_2001} used a similar concept of input file templates for parameter sweeping. Besides not being publicly available, this tool was less general than parasweep in the types of sweeps and the schedulers supported. The more recent preconfig \cite{nedelec_preconfig_2017} is a tool for generating configuration files, but does not handle dispatching or post-processing. Tools such as GNU Parallel \cite{tange_gnu_2011} and Slurm or PBS job arrays, while not designed solely for parameter sweeps, are sometimes used to facilitate them by automating the process of running the simulations with the different parameter values in parallel. However, these tools require the parameters to be passed through command-line arguments, which necessitates parsing within the model software. Moreover, those relying on job arrays only work with their respective schedulers. None of the tools in the latter group supports different types of parameter sweeps, keeps records of the parameters used, or facilitates post-processing. Thus parasweep, unlike previous tools, provides a complete cross-platform solution for generating, dispatching in parallel, and post-processing parameter sweeps, relying on a simple template-based system.

Throughout the paper, we refer to the program on which we run a parameter sweep as the \emph{model}, a particular assignment of values to each of the parameters as a \emph{parameter set}, a single run of the model with a particular parameter set as a \emph{simulation}, and the collection of all the simulations as the \emph{sweep}.

\section{Software description}

\subsection{Software architecture}
parasweep is written in Python, a cross-platform, general-purpose language widely used for scientific applications. Although Python is an interpreted language and generally slower than compiled languages such as C or C++, this is not likely to be a bottleneck since the time for generating parameter sets and filling out a template is insignificant compared to the simulation time for the vast majority of applications. An object-oriented structure makes the sweep type, dispatching, template engine, and generation of simulation identifiers entirely modular, allowing parasweep to be easily extensible. All features are documented and tested with a test suite.

The idea of parasweep is to leverage the existing configuration files of the given model. These files have a single value for each parameter, but parasweep allows parameter values to be swept over with little effort. This is done by providing parasweep with a \emph{configuration file template}, which is identical to the configuration file, except with placeholders where the parameters to be swept over will be inserted. The user specifies the parameter sweep, which produces sets of parameters to be given to the model. Using the template, parasweep generates a configuration file for each parameter set, and assigns this set of parameter values a unique identifier (the simulation ID). (This is explained in more detail below.) The only modification that needs to be made to the model is to receive the simulation ID as a command-line argument, read the generated configuration file corresponding to that ID, and write the output to a file also corresponding to that ID. This approach thus requires no major changes to the configuration system of the model, no parsing of parameter values through the command-line (which can be time-consuming and must be modified for every parameter added), and is easy to set up in whatever language the model is written. The simulation ID provides a way to associate the parameter set with the output for every simulation in the sweep.

The basic sequence for running parameter sweeps with parasweep is:
\begin{enumerate}
	\item Generate the sets of parameter values.
	\item For each set of parameter values:
	\begin{enumerate}
		\item Assign a simulation ID.
		\item Using the configuration template, fill in the parameter values into a configuration file with the simulation ID in the name.
		\item Dispatch a simulation with the simulation ID as a command-line argument.
		\item In the model program, open the configuration file with the given simulation ID, read the parameters, and run the simulation. Output to a file corresponding to the same ID.
	\end{enumerate}
	\item Return a mapping between the sets of parameter values and the simulation IDs.
\end{enumerate}
The sweep type, assignment of simulation IDs, template engine, dispatching, and mapping type are all configurable and several options are provided for each within parasweep. We discuss the options for sweep types, dispatching, and mapping below. As mentioned above, parasweep's modular structure makes it easy to extend.

\subsection{Software functionalities}
The implemented sweeps are Cartesian product sweeps, filtered Cartesian product sweeps, set sweeps, and random sweeps. In Cartesian product sweeps (sometimes known as grid sweeps), all the possible combinations of the given parameter values are run. Filtered Cartesian product sweeps allow the user to specify in addition a filtering function of the parameters, and only those parameter sets that meet the condition of the filter are run. This can be used, for example, to run a parameter sweep of a model that takes parameters $x$ and $y$, but with the condition that $x > y$. Set sweeps run only the parameter sets specified by the user. Random sweeps sample each variable as an independent probability distribution, with a wide variety of distributions from which to select.

Simulations can be dispatched by spawning processes locally, a useful option for multi-core computers. Alternatively, a large number of job schedulers typically found on high-performance computing (HPC) systems, both cluster and grid, are supported using the Distributed Resource Management Application API \parencite[DRMAA:][]{troger_standardization_2007} if it is installed on the system. This includes Slurm and PBS/Torque among a number of others.

For post-processing, parasweep keeps track of the simulation IDs assigned to each parameter set. For a Cartesian sweep, this mapping can be naturally represented as an $n$-dimensional array, where $n$ is the number of parameters in the sweep. The mapping for Cartesian sweeps is thus a labelled array provided by xarray, a powerful library for handling multidimensional labelled data \parencite{hoyer_xarray_2017}. This array can be saved to disk as a netCDF file for future reference. For the other types of sweeps, since a multidimensional array is not a parsimonious representation, the mapping is a dictionary (hash map) between the simulation IDs and the parameter sets used. This can be saved to disk as a JSON file.

\section{Illustrative example}

We present the following example of the usage of parasweep. More examples, showing all the major features of parasweep, are available in the documentation.

\subsection{The model}
Our model in this case is a Fortran program \verb|lorenz|, which simulates the Lorenz '63 model of convection \cite{lorenz_deterministic_1963} and outputs its largest Lyapunov exponent. The Lorenz model takes three parameters, $\beta$, $\sigma$, and $\rho$, and it is known that it is chaotic (exhibits sensitive dependence on initial conditions) for some values of these parameters and not for others. We wish to know for which parameter sets it is chaotic, and we can determine this by checking whether the largest Lyapunov exponent of the system is positive. The definition and algorithm\footnote{The algorithm tracks two points close to each other on the attractor and rescales the vector that connects them \cite{sprott_chaos_2003}.} for computing the largest Lyapunov exponent is not important for our purposes. The full code for this example is provided in the parasweep code repository, but in this section we discuss only the necessary changes to be able to conduct parameter sweeps with it.

The model reads a configuration file \verb|params.nml| which contains the values of $\beta$, $\sigma$, and $\rho$; we now modify it to instead use the file \verb|params_{sim_id}.nml|, where the simulation ID \verb|sim_id| is provided as a command-line argument.

It suffices to change
\begin{minted}[frame=single]{fortran}
namelist /params/ beta, sigma, rho

open(1, file="params.nml")
read(1, nml=params)
\end{minted}
to
\begin{minted}[frame=single]{fortran}
namelist /params/ beta, sigma, rho
character(30) :: sim_id

call get_command_argument(1, sim_id)
open(1, file="params_" // trim(sim_id) // ".nml")
read(1, nml=params)
\end{minted}

We also modify the model to output to the filename \verb|results_{sim_id}.txt| instead of \verb|results.txt|. We change
\begin{minted}[frame=single]{fortran}
open(2, file="results.txt", action="write")
write(2, *) lyap
\end{minted}
to
\begin{minted}[frame=single]{fortran}
open(2, file="results_" // trim(sim_id) // ".txt", action="write")
write(2, *) lyap
\end{minted}

\subsection{The configuration template}
Suppose the \verb|options.txt| looked like the following:
\begin{minted}[frame=single]{text}
&params
beta = 2.67,
sigma = 10,
rho = 28
/
\end{minted}
Here $\beta$, $\sigma$, and $\rho$ are hard-coded. To make the parameters able to be swept over, we simply need to indicate where they must go and give them an identifier surrounded by curly braces:
\begin{minted}[frame=single]{text}
&params
beta = {beta},
sigma = {sigma},
rho = {rho}
/
\end{minted}
This is the template, into which the parameter values will be substituted for every simulation in the sweep. We save it as \verb|template.txt|. Note that this is the format of the configuration file for this particular model, and a different template has to be created for every model in order to run a parameter sweep on it.

\subsection{The command}\label{sec:command}
We can now run a parameter sweep. Suppose we want to try 3 evenly spaced values of $\beta$ between 2 and 4, 10 values of $\sigma$ between 2 and 20, and 10 values of $\rho$ between 2 and 30. Then the sweep can be run as follows:
\begin{minted}[frame=single]{python}
import numpy
import xarray

from parasweep import run_sweep, CartesianSweep

sweep_params = {'beta': numpy.linspace(2, 4, 3),
                'sigma': numpy.linspace(2, 20, 10),
                'rho': numpy.linspace(2, 30, 10)}

sweep = CartesianSweep(sweep_params)

mapping = run_sweep(command='./lorenz {sim_id}',
                    configs=['params_{sim_id}.nml'],
                    templates=['template.txt'],
                    sweep=sweep)
\end{minted}
This means the following:
\begin{itemize}
	\item \verb|command|: specifies the command to run a simulation with the model. Note that \verb|{sim_id}| indicates where the simulation ID for each simulation in the sweep is to be substituted in the command; \verb|sim_id| is a special keyword that must be used in both the \verb|command| and the \verb|configs| arguments.
	\item \verb|configs|: sets the name of the configuration file that will be created for each simulation in the sweep, where \verb|{sim_id}| indicates where the simulation ID is to be substituted in the filename.
	\item \verb|templates|: specifies the location of the configuration file template.
	\item \verb|sweep|: specifies the sweep type. In this case, we select a Cartesian product sweep and provide the parameter values for each parameter we would like to sweep over. Since there are 3 possible values of $\beta$, 10 possible values of $\sigma$, and 10 possible values of $\rho$, 300 simulations will be run.
\end{itemize}
These are the required arguments to \verb|run_sweep|. Descriptions of all the arguments is available in the documentation.

\subsection{Post-processing}
\begin{figure}
	\centering
	\includegraphics[scale=0.4]{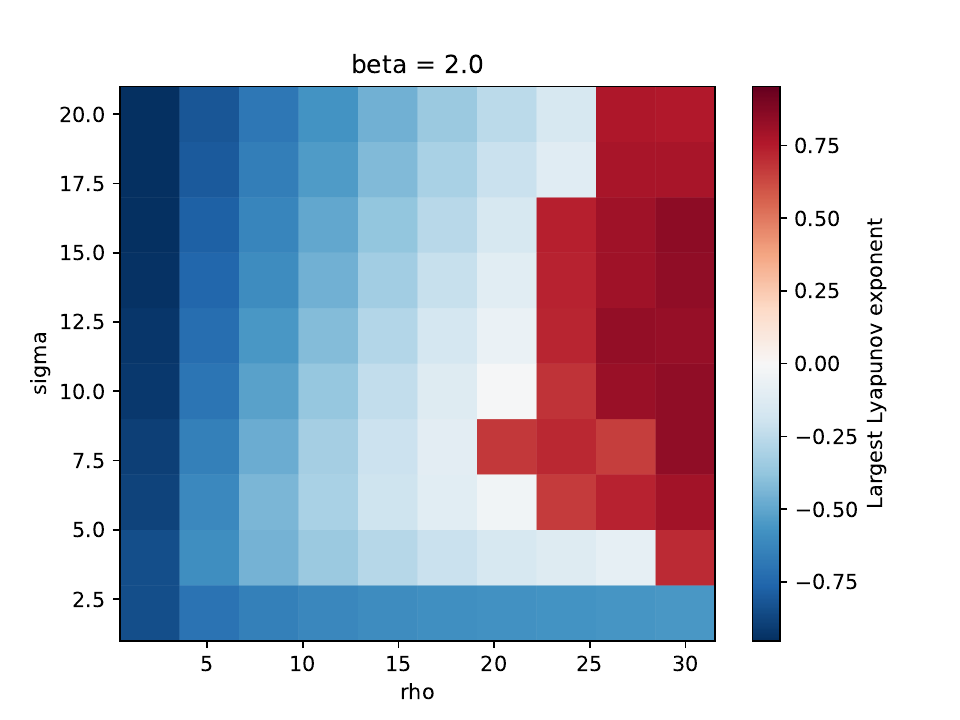}
	\caption{The largest Lyapunov exponent of the Lorenz model as a function of $\rho$ and $\sigma$, with fixed $\beta=2$.}
	\label{fig:lyap}
\end{figure}

We now want to extract the results of the simulations and plot them. We use the \verb|mapping| object returned after calling the \verb|run_sweep| function. It is an xarray \verb|DataArray| object, a labelled N-dimensional array. The coordinates are the sweep parameters and the ``data'' is the simulation IDs. This makes it easy for programs to retrieve the simulation output by the parameter values rather than having to specify the simulation IDs manually. The example below, executed after the code in section \ref{sec:command}, selects the first $\beta$ (in this case, $\beta=2$) and plots the largest Lyapunov exponent as a function of $\rho$ and $\sigma$.
\begin{minted}[frame=single]{python}
def get_output(sim_id):
    filename = f'results_{sim_id}.txt'
    return numpy.loadtxt(filename)


lyap = xarray.apply_ufunc(get_output, mapping, vectorize=True)
lyap = lyap.rename('Largest Lyapunov exponent')

lyap.isel(beta=0).plot()
\end{minted}
This will produce Figure \ref{fig:lyap}. The chaotic regime of the parameter space can then be easily read off as those parameter sets which result in a positive largest Lyapunov exponent (the red regions of the plot). This is just one example of the types of post-processing that can be done.

\section{Impact}
parasweep considerably simplifies the process of running parallel parameter sweeps, with applications to many scientific fields. As of January 2021, parasweep has had over 8500 downloads from PyPi (the official Python package repository) alone, not counting downloads from GitHub, which are not tracked. The author is aware of parasweep being used for running parameter sweeps of a coupled atmosphere--ocean model, a mathematical model of epithelial cells, electronic circuit simulations, and an ensemble forecasting method for dynamical systems \cite{bach_ensemble_2021}.

\section{Conclusions}
\label{}

We present parasweep, a Python utility for generating, dispatching, and post-processing of parameter sweeps. parasweep allows for easy generation of parameter sweeps with existing models by using a template-based system. We discuss its potential to be useful in a wide variety of scientific applications, and present an illustrative example.

Although designed for parameter sweeps, parasweep can be useful for any application that requires generation of configuration files, dispatching tasks in parallel, and post-processing. The sweep type, assignment of simulation IDs, template engine, dispatching, and mapping type are all modular within parasweep, making it easily extensible beyond its current capabilities.

\section*{Declaration of competing interest}
The authors declare that they have no known competing financial interests or personal relationships that could have appeared to influence the work reported in this paper.

\section*{Acknowledgments}
\label{}

The author thanks Eugenia Kalnay for helpful comments. The author acknowledges the University of Maryland supercomputing resources (\url{http://hpcc.umd.edu}) made available for conducting the research reported in this paper.

\printbibliography

\end{document}